\documentstyle[prd,aps,floats]{revtex}
\begin{document}
\draft

\input epsf \renewcommand{\topfraction}{0.8}
\twocolumn[\hsize\textwidth\columnwidth\hsize\csname
@twocolumnfalse\endcsname


\title{A model for fluctuating inflaton coupling: (s)neutrino induced 
adiabatic perturbations and non-thermal leptogenesis}

\author{Anupam Mazumdar}
\address{CHEP, McGill University, Montr\'eal, QC, H3A~2T8, Canada.}
\maketitle
                      
\begin{abstract}
We discuss an unique possibility of generating adiabatic density 
perturbations and leptogenesis from the spatial fluctuations of the 
inflaton decay rate. The key assumption is that the initial 
isocurvature perturbations are created in the right handed 
sneutrino sector during inflation which is then converted into 
adiabatic perturbations when the inflaton decays. We discuss distinct 
imprints on the cosmic micro wave background radiation, which can 
distinguish non-thermal versus thermal leptogenesis.
\end{abstract}

\vskip2pc]

\renewcommand{\thefootnote}{\arabic{footnote}}
\setcounter{footnote}{0}


Inflation is the main contender for explaining the observed adiabatic
density perturbations with a nearly scale invariant
spectrum~\cite{Linde}. However, recently various alternative mechanisms
for generating adiabatic density perturbations have been
discussed, particularly converting the isocurvature perturbations of
some light field into the adiabatic perturbations in the
post-inflationary Universe~\cite{Enqvist02,Enqvist03,Sloth}.  Another
interesting proposal is that the perturbations could be generated from
the fluctuations of the inflaton coupling to the Standard Model
degrees of freedom~\cite{Dvali,Kofman}. It has been argued that the
inflaton coupling strength to the ordinary matter, instead of being a
constant, could depend on the vacuum expectation values (VEV) of
various fields in the theory. These fields are none other than the
flat directions of the minimal supersymmetric standard model (MSSM).
However the authors in \cite{Dvali} treated the flat directions
without considering the fact that the flat directions are lifted at a
non-renormalizable level. In~\cite{Mazumdar}, the authors have
demonstrated the importance of non-renormalizable potential terms for
the flat directions, which leads to dramatic changes in the estimation
of the amplitude of the density perturbations in the original
scheme. The amplitude of the perturbations dampens after the end of
inflation, because the flat directions evolve after the end of
inflation until the decay of the inflaton. The damping of the
amplitude of the perturbations acts as a main challenge for realizing
such a novel scheme (for a review on MSSM flat direction and
cosmology, see~\cite{Enqvist}).

The idea is that if the MSSM condensates (made up of squarks and
sleptons) are light during inflation then their quantum fluctuations
can give rise to spatial fluctuations in the inflaton coupling
strength. When the inflaton decays, the adiabatic density
perturbations are created because the isocurvature perturbations
generated by the flat direction during inflation is transferred to the
adiabatic ones right at the time of decay.  If the flat direction
evaporates into baryons, it will give rise to the baryon isocurvature
fluctuations, which can be constrained from cosmic microwave
background radiation (CMB).

A particularly interesting implementation of this scenario is to
consider the right handed (s)neutrinos (supersymmetry guarantees the
presence of the right handed sneutrino). If the right handed (s)neutrino
is Majorana, then it provides a natural explaining for the observed
small neutrino masses via see-saw mechanism, $m_\nu
\approx\left(m^{2}_{\rm D}/M_N\right)$~\cite{seesaw}, where $m_{\rm
D}$ is the Dirac mass obtained from the Higgs VEV.

The right handed (s)neutrino is also a source for $L$ and/or $B-L$
violation. Therefore it can be responsible for the observed baryon
asymmetry. Leptogenesis requires $L$ or $B-L$ violating interactions,
$C$ and $CP$ asymmetry, and an out of equilibrium condition. The first
two are well served by the right handed (s)neutrinos, and the last
condition naturally arises in any inflationary cosmology.

Now let us briefly discuss the status of inflation. Undoubtfully,
inflation is the most natural mechanism which makes the Universe
homogeneous, flat, and isotropic. A single field slow roll inflation
is also the most beautiful way of explaining adiabatic density
perturbations. However until now there is hardly any connection
between the inflationary sector and the known particle physics
sector. In most of the cases the inflaton is a gauge singlet, which
leads to some degree of speculation on how the inflaton couples
to the SM gauge particles~? what is the inflaton potential~?,
etc. The coupling of the inflaton to the SM fields is essential if the
inflationary paradigm wishes to make connection to the hot big bang
cosmology.

In this paper our aim is to present a simple toy model where we
illustrate two important aspects. The first one is to consider that
the sneutrino induced isocurvature fluctuations which can generate adiabatic
perturbations through inflaton decay. Second, we will show that in a
non-thermal leptogenesis scenario the produced baryon-isocurvature
fluctuations can be testable from CMB.

In this regard it is natural to come up with a scenario where the
inflaton couples only through the right handed (s)neutrino sector.
Note that the right handed (s)neutrino acts as a mediator which
connects the two disparate sectors, e.g., the inflaton and the SM via
lepton (sleptons) and Higgs (Higgsinos).  Therefore such a model is
not only economical in terms of achieving density perturbations and
lepton asymmetry, but also providing us with the correct relativistic
species required for the nucleosynthesis.

For the purpose of illustration, let us consider a very simple toy model,
\begin{equation}
\label{super} 
W \supset {1 \over 2} g \Phi {\bf N}{\bf
N} + h{\bf N} {\bf H}_u {\bf L} + {1 \over 2} M_N {\bf N} {\bf N}\,,
\end{equation}
where $\Phi$, ${\bf N}$, ${\bf L}$, and ${\bf H}$, respectively 
stand for the inflaton, the right handed neutrino, the lepton doublet, 
and the Higgs (which gives mass to the top quark) superfields. Also 
$M_N$ denotes right handed (s)neutrino masses, and $g,~h$ correspond to 
the Yukawas. Being a SM gauge singlet the inflaton can naturally
couple to the right handed neutrino sector with a renormalizable and  
a non-renormalizable coupling
\begin{equation}
\label{coupling}
g= g_{0}\left(1+\frac{{\bf N}}{M_{\rm p}}+...\right)\,,
\end{equation}
where we assume that the non-renormalizable scale is the Planck mass,
$M_{\rm p}=2.4\times 10^{18}$~GeV. For simplicity we have omitted all
the indices in $h$ matrix and superfields, and we work on a basis
where the Majorana mass matrix is diagonal. Further simplifications
can be made for almost degenerate right handed (s)neutrinos where
$M_N$ is essentially the same for all.  It is also conceivable in this
case that the inflaton is coupled with the same strength to three
right handed (s)neutrinos with a mass hierarchy $M_{N}\gg m_{\phi}$,
where $m_{\phi}$ can be treated as the mass of the inflaton in the
minimum of its potential. Note that this is just a working example of
non-thermal leptogenesis. We will highlight why we stress upon
non-thermal leptogenesis compared to thermal leptogenesis.  More
complicated scenarios on non-thermal leptogenesis can be
constructed~\cite{Allahverdi1,Berezhiani}.

The inflaton sector is still unknown, except that it is responsible
for driving inflation, which could be for e.g., brane driven
inflation, fast rolling inflation, kinetic driven inflation, assisted
inflation, false vacuum inflation, etc. We further assume that the
inflaton decays perturbatively. At this point one might suspect that
the above coupling,~$g$, in Eq.~(\ref{super}), would give rise to a
large mass contribution to the sneutrino through the inflaton vev, and
the sneutrino would simply roll down to the bottom of the
potential. However note that the inflaton vev need not be always large 
in order to inflate, for instance in a false vacuum inflation 
the inflaton could be fairly close to the bottom of its own potential. 
Nevertheless, let us imagine that we are working in a regime
where the coupling, $g$, is such that the effective mass for the
sneutrino is less than the Hubble expansion, $m_{eff,~\phi}<
M_{eff,~N}\lesssim H$, where $m_{eff,\phi}$ is the effective mass of
inflaton during inflation. We will comment on the situation when the
sneutrino masses are heavy compared to the Hubble expansion during
inflation. From here onwards we remove the subscript, ${eff}$.

An important point to note here is that during inflation the 
quantum fluctuations are created in the sneutrino sector, and the 
perturbations in the inflaton sector is assumed to be negligible, 
therefore, the perturbations arises only from the known particle physics 
sneutrino sector. The perturbations on a comoving scale larger 
than the Hubble scale can be foliated in terms of the curvature 
perturbations on a  finite energy density surface: 
$ds^2=a^2(t)\left(1+2\zeta\right)dx^{i}dx^{j}$, where $\zeta$ is 
a metric perturbation written in a proper coordinate system.
In presence of more than one scalar fields the total curvature 
perturbations $\zeta$ evolves outside the horizon due to non-vanishing
pressure perturbations~\cite{Brandenberger}.
\begin{equation}
\label{curv}
\dot\zeta=-\frac{H}{\rho+P}\delta P\,.
\end{equation}
where $\rho,~P$ are the energy density and the pressure. For a single
field inflation, $\zeta={\rm constant}$, but in a multi-field case 
$\delta P$ is a non-zero quantity due to the entropy perturbations, which 
can be defined in our case as
\begin{equation}
\label{entropy}
S_{\phi,\widetilde N}=3\left(\zeta_{\phi}-\zeta_{\widetilde N}\right)=-3H\left(
\frac{\delta \rho_{\phi}}{\dot\rho_{\phi}}-\frac{\delta\rho_{\widetilde N}}
{\dot\rho_{\widetilde N}}\right)\,.
\end{equation} 
Over-dot denotes differentiation w.r.t. coordinate time. 
Following our assumption the {\it initial} entropy 
perturbation becomes $S_{\phi,\widetilde N}\sim -3\zeta_{\widetilde N}$. 
For the Gaussian perturbations, we obtain
\begin{equation}
\label{p}
{\cal P}_{\widetilde N}^{1/2}=\frac{H_{\ast}}{2\pi}\,,
\end{equation}
where $\ast$ denotes when the interesting perturbations leave the
horizon, $k=a_{\ast}H_{\ast}$. Note that the entropy perturbations
feed the total curvature perturbations, therefore the entropy
perturbations along with the individual perturbations,
$\zeta_{\phi},\zeta_{\widetilde N}$, evolve in time. Though we do not
prove this here, but intuitively we can see that in order to obtain
the adiabatic density perturbations, the total curvature perturbation
must become constant outside the horizon at the time of inflaton
decay, when $\left.\zeta_{\phi}\right|_{decay}\sim \zeta_{\widetilde
N}$, and therefore, $\zeta$, becomes constant on large scales. Thus
converting its initial isocurvature fluctuations into the adiabatic
ones.

Note that besides the fluctuations in the sneutrino sector, there will
be fluctuations in the Higgses and the sleptons also. However their
perturbations will not account for the baryon isocurvature fluctuations
in the above set up, for the time being we will neglect them.

The spectral index for the perturbations can be written as
\begin{equation}
n_{\zeta}-1\equiv \frac{d\ln {\cal P}_{\widetilde N}}{d\ln k}=
2\frac{\dot H_{\ast}}{H_{\ast}^2}+\frac{2}{3}
\frac{M_{N}^2}{H^2_{\ast}}\,.
\end{equation}
Therefore as long as the Hubble expansion is slowly varying, 
and $M_{N}\leq H$, we can obtain a scale invariant density 
perturbation.

Now let us study the decay of the inflaton. The main decay 
mode of the inflaton is four-body final states consisting of two 
Higgs/Higgsino-lepton/slepton particles (and their $CP$ transforms),
this is due to the fact that the inflaton is decaying via off-shell
(s)neutrino. The effective superpotential after integrating out ${\bf N}$,
is given by
\begin{equation} 
\label{effective}
W_{eff} \supset {1 \over 2 M^{2}_{N}}gh^2 \Phi 
({\bf H}_u {\bf L})({\bf H}_u {\bf L})\,. 
\end{equation}
For simplicity we may consider a situation when 
the (s)neutrinos are almost degenerate, e.g. 
$\Delta M_{N}<M_{N}$, and the Yukawa texture is 
such that the diagonal entries (h) and off-diagonal 
entries ($h^{\prime}$) follow $h^{\prime}<h$.

There are total nine final states, seven of them consist of two
fermions and two scalars, and there are also two final states
consisting of four scalars. Summing up all the final states, the 
decay rate and the final reheat temperature are given by
\begin{equation}
\label{decay}
\Gamma_{\rm d} \simeq \frac{21 g^2h^4 m_{\phi}^5}{2^{14} \pi^5 M_{N}^4}\,,~~~
{T_{\rm R} \over m_\phi} \simeq \frac{10^{-7/2}g h^2 m^{3/2}_{\phi} 
M^{1/2}_{\rm P}}{M^{2}_{N}}\,. 
\end{equation}
However note that the reheat temperature obtains a spatial 
fluctuations due to Eq.~(\ref{coupling}),
\begin{equation}
\label{rel}
\frac{\delta T_{\rm R}}{T_{\rm R}}=-\frac{1}{3}\frac{\delta g}{g}\sim 
-\frac{\delta \widetilde N}{3M_{\rm p}}\sim -\frac{H_{\ast}}{6\pi M_{\rm p}}\,.
\end{equation}
The factor $-1/3$ arises because during the decay of the inflaton the 
average energy density goes as $\rho\sim a^{-3}$, see for details 
\cite{Dvali,Mazumdar}. For the Gaussian perturbations, $\widetilde N\gg H$, 
and following Eq.~(\ref{p}) we obtain the right amplitude for the
density perturbations provided $H_{\ast}\sim 10^{-5}M_{\rm p}$. Note
that there is no damping in the sneutrino fluctuations after the end
of inflation in this case. Further note that if $\widetilde N\ll H$,
then the Gaussian amplitude of the perturbations will be
damped~\cite{Liddle}, which we would like to avoid in this example.

Let us now obtain the lepton asymmetry in this model. The
$CP$ asymmetry is obtained through the inference between the tree level 
and the one loop (vertex and self energy) corrected diagrams,
Net $CP$ asymmetry in the off-shell case is quite different compared 
to the on-shell leptogenesis~\cite{Allahverdi1}. The self energy correction 
comes out to be twice as much as the vertex correction for 
$m_{\phi}\ll M_{N}$. Final $CP$ asymmetry is then given by 
\begin{equation} 
\label{asymmetry1}
\epsilon_{CP} \simeq  - {3 \over 8 \pi} \times {\sum_{i,n,l} {{\rm Im} 
\left[({\bf h} {\bf h}^{\dagger})_{ni} 
({\bf h} {\bf h}^{\dagger})_{nl} ({\bf h} {\bf h}^{\dagger})_{il}\right]
m^{2}_{\phi}
\over M^{3}_{i} M^{2}_{n} M_{l}} \over \sum_{i,n} 
{([{\bf h} {\bf h}^{\dagger}]_{in})^{2} \over M^{2}_{i} M^{2}_{n}}}\,,
\end{equation}              
where $i,n,l=1,2,3$. The produced lepton asymmetry reads as
\begin{equation} 
\label{finallepton}
{n_{L} \over n_{\phi}} \simeq {3 \over \pi} {\delta h h'^{2} \over h} 
{{\Delta M}_{N} \over M_{N}} \left({m_\phi \over M_N}\right)^{2}\,,
\end{equation}
where $\delta h$ is nearly equal diagonal entries of the Yukawa matrix.
For nearly degenerate case $\delta h/h \sim \Delta M_{N}/2M_{N}$.

The total asymmetry in the baryons (after taking into account of the
sphaleron effects) can be expressed as  
\begin{eqnarray} 
\label{final}
\eta_{\rm B} &= &\left(\frac{n_{B}}{n_{\phi}}\right)\left(
\frac{n_{\phi}}{s}\right)\, \nonumber \\
&\simeq &{1 \over \pi} {{\delta h} h'^{2} \over
h^3} {\Delta M_N \over M_N} \left({M_N m_{\nu} \over {\langle
H^{0}_{u}\rangle}^2}\right)
\times \left ({m_\phi \over M_N}\right )^2\left({T_{\rm R} \over
m_\phi}\right)\,,
\end{eqnarray}
where $s=(2\pi^2/45)g_{\ast}T_{\rm R}^3$. Here
$n_{\phi}/s$ denotes the dilution from reheating. By using
the expression for the reheat temperature and 
$m_\nu \simeq (h^2 \langle H^{0}_{u}\rangle^2/ M_N)$, we 
finally obtain
\begin{equation}
\label{result}
\eta_{\rm B} \simeq 4.10^{-49/2} g {{\delta h} h'^{2} \over
h^3} {\Delta M_N \over M_N} {m^{7/2}_{\phi}
M^{1/2}_{\rm
P} \over M^{2}_{N} {\langle H^{0}_{u} \rangle}^4}~~(1 {\rm GeV})^2\,,
\end{equation}        
where we have considered $m_\nu \approx 0.1$~eV, and $\langle
H^{0}_{u} \rangle = 174$~GeV. If we demand degenerate light neutrino
masses, then we can further simplify, Eq.~(\ref{result}),~~
$\eta_{\rm B}\simeq$
\begin{equation}
\label{finalresult}
\frac{2.10^{-49/2} g h^{\prime 2}}{
h^2} \left({\Delta M_N \over M_N}\right)^2 {m^{7/2}_{\phi}
M^{1/2}_{\rm
P} \over M^{2}_{N} {\langle H^{0}_{u} \rangle}^4}~~(1 {\rm GeV})^2\,. 
\end{equation}
Some numerical examples for nearly degenerate heavy right handed
(s)neutrinos, with $M_N = 10 m_\phi$ and $10^{-1} \leq h'/h \leq 1$,
we obtain the desired baryon asymmetry for $10^{-3}\leq g \leq 1$ and
$10^{11}~{\rm GeV} \leq m_\phi \leq 10^{13}$~GeV, which result in
reheat temperature: $10^6~{\rm GeV} \leq T_{\rm R} \leq 10^8$~GeV. At
this point one might worry upon the coupling strength, $g$, because
the inflaton picks up an effective mass term, $g\langle {\widetilde
N}\rangle$, which has to be smaller than the inflaton mass,
$m_{\phi}$, arising solely from the inflaton sector, in order to keep
the successes of slow roll inflation. On the other hand in order to
generate the Gaussian fluctuations, ${\widetilde N}\gg H_{\ast} \sim
10^{-5}M_{p}$, see Eq.~(\ref{rel}). This leads to $g\ll 1$, however,
an exact magnitude of $g$ will depend on a particular inflationary
model.  Further note that the reheat temperature is well below thermal
and non-thermal gravitino over-production~\cite{Ellis84,Maroto}.

The most important point is to note that the baryon asymmetry is 
proportional to $g$, see our final result, Eq.~(\ref{finalresult}). 
Therefore baryons also feel the spatial fluctuations.
\begin{equation}
\label{excite}
\frac{\delta \eta_{B}}{\eta_{B}}\sim -\frac{1}{3}\frac{\delta g}{g}\sim 
-\frac{\delta\widetilde N}{3M_{\rm p}}\sim 
\frac{\delta T_{\rm R}}{T_{\rm R}}\neq 0\,.
\end{equation}
The origin of $-1/3$ factor has the same origin as in
Eq.~(\ref{rel}). Note that the fluctuations in the baryon asymmetry is
proportional to the fluctuations in the inflaton coupling, and
therefore fluctuations in the reheat temperature.  This shows that the
baryonic asymmetry does not follow honest to God the adiabatic density
perturbations, instead {\it perturbation in baryons is correlated
baryon-isocurvature in nature}.

The baryon-isocurvature fluctuation leaves its imprint on cosmic
micro wave background radiation. Moreover the fluctuations are
correlated. The WMAP data provides mild constraint on the
correlated-cold dark matter-isocurvature fluctuations~\cite{WMAP},
which can be translated in terms of the baryon isocurvature
fluctuations as $|S_{B}/\zeta| < 0.32(\Omega_{CDM}/\Omega_{B})\sim
1.85$ at $95\%$ confidence level, where $S_{B}$ is the
baryon-isocurvature fluctuations. In our toy model
$S_{B}=\delta\eta_{B}/\eta_{B}=\delta T_{\rm R}/T_{\rm R}$.  In
particular
$\zeta=-H\delta\rho/\dot\rho=(1/4)\delta\rho_{\gamma}/\rho_{\gamma}=\delta
T_{\rm R}/T_{\rm R}$, where the subscript $\gamma$ denotes
radiation. Therefore we find $|S_{B}/\zeta|=1$ in our case, which is
well within the WMAP constraint on the baryon-isocurvature perturbations.

We point out here that the above feature of correlated
baryon-isocurvature perturbations is only present in a non-thermal case,
this is the reason why we pursued on non-thermal leptogenesis.  In
non-thermal leptogenesis there is an explicit dependence on the reheat
temperature, see Eq.~(\ref{final}).  This is indeed an interesting
feature of a non-thermal leptogenesis which is absent in a thermal
case. In a thermal leptogenesis the net asymmetry is proportional to a
$CP$ asymmetry and not to a temperature~\footnote{In general
thermal/non-thermal leptogenesis provides the net asymmetry as
$\eta_{B/L}\sim
\epsilon_{CP} \times f(T_{d}/M)$, where $T_{D}$ is the temperature of
the decaying particle, e.g. the right handed neutrino. In case of
thermal leptogenesis $T_{d}\sim M$, and there is no temperature
dependence. However in a non-thermal case $T_{d}\neq M$.}. We find
that the constraints on the baryon-isocurvature perturbations can act
as a tool for differentiating thermal versus non-thermal leptogenesis
mechanisms.

Before we conclude our paper, we comment on couple of interesting
points. In an opposite limit when $m_{\phi}>M_{N}$, the inflaton
decays via on-shell right handed (s)neutrino to the SM leptons and
Higgs. This case is even better because during inflation the condition
$\widetilde N \gg H_{\ast}$ is satisfied even better, because the
sneutrino is lighter, $M_{N}\lesssim m_{\phi}\ll H$. However one has
to ensure that a thermal regeneration of the baryon asymmetry is really
small. Finally we comment on a heavy right handed neutrino masses
compared to the Hubble expansion, in this case the sneutrino
perturbations will be $\chi^2$ in nature, and usually the amplitude of
the perturbations comes out to be small~\cite{Liddle}. Nevertheless,
they can also provide interesting imprints on CMB through the tilt in
the spectral index. We leave this for future investigation.

In fact we could also imagine perturbing the other Yukawa coupling,
$h$, similar to Eq.~(\ref{coupling}), and the inflaton mass due to the
fluctuations in the sneutrino vev.  By inspecting the reheat
temperature, Eq.~(\ref{decay}), and the baryon asymmetry,
Eq.~(\ref{final}), we obtain $\delta T_{\rm R}/T_{\rm R}=-(2/3)(\delta
h/h)=\delta\eta_{\rm B}/\eta_{\rm B}$, assuming that the fluctuations
are arising only from the diagonal elements of the Yukawas. In this
case the prediction on the baryon-isocurvature fluctuations remains,
$|S_{\rm B}/\zeta|=1$. However the fluctuating inflaton mass gives
rise to $|S_{\rm B}/\zeta|=1.4$, which is still within the WMAP limit
on $|S_{\rm B}/\zeta|<1.85$~@ $95~\%$ confidence level~\cite{WMAP}.

As a final remark, there could be other sources for the isocurvature 
perturbations during inflation, including the most competitive candidate 
``cold dark matter''. However within supersymmetry excellent conditions arise 
naturally for their thermal production.

To summarize our paper, we point out that any supersymmetric
leptogenesis scenario is a potential candidate for succeeding in
generating the adiabatic density perturbations from the sneutrino fluctuations
during inflation and generating the baryon asymmetry. The nature of 
perturbations (Gaussian/non-Gaussian) certainly depends on the 
mass scales, e.g., for $M_{N}\ll H_{\ast}$, and $\widetilde N \gg H_{\ast}$ 
the perturbations are Gaussian. This may not be the case if $M_{N}=3H$ during
inflation, see~\cite{Liddle}. Note that our model is economical because
it achieves several goals at a time. Finally we have found 
a very important bench mark which can potentially differentiate 
thermal versus non-thermal leptogenesis from CMB.

\vskip10pt

A. M. is a CITA National fellow. The author has benefited from the
discussion with Rouzbeh Allahverdi, Robert Brandenberger, Zurab
Berezhiani, Guy Moore, David Wands, and valuable comments from the
anonymous referees.





\begin{references}

\vspace*{-1.8truecm}

\bibitem{Linde}
For a review, see: A.D. Linde, {\it Particle Physics And Inflationary
Cosmology}, Harwood (1990).

\bibitem{Enqvist02}
K.~Enqvist, S.~Kasuya and A.~Mazumdar,
Phys.\ Rev.\ Lett.\  {\bf 90}, 091302 (2003)

\bibitem{Enqvist03}
K.~Enqvist, A.~Jokinen, S.~Kasuya and A.~Mazumdar,
hep-ph/0303165.

\bibitem{Sloth}
K.~Enqvist and M.~S.~Sloth, Nucl.\ Phys.\ B {\bf 626}, 395 (2002);
D.~H.~Lyth and D.~Wands, Phys.\ Lett.\ B {\bf 524}, 5 (2002).
T.~Moroi and T.~Takahashi, Phys.\ Lett.\ B {\bf 522}, 215 (2001)
[Erratum-ibid.\ B {\bf 539}, 303 (2002)];
S. Mollerach, Phys. Rev. D {\bf 42}, 313 (1990).
M.~Postma, Phys.\ Rev.\ D {\bf 67}, 063518 (2003);


\bibitem{Dvali}
G. Dvali, A. Gruzinov, and M. Zaldarriaga, astro-ph/0303591.

\bibitem{Kofman}
L. Kofman, astro-ph/0303614.

\bibitem{Mazumdar}
K.~Enqvist, A.~Mazumdar and M.~Postma,
Phys.\ Rev.\ D {\bf 67}, 121303 (2003).

\bibitem{Enqvist}
For a review, see K.~Enqvist and A.~Mazumdar, Phys.\ Rept.\  {\bf 380}, 
99 (2003).



\bibitem{seesaw}
M. Gell-Mann, P. Ramond, and R. Slansky, in {\it Supergravity},
eds. P. van Niewenhuizen and D.Z. Freedman (North Holland 1979);
T. Yanagida, Proceedings of {\it Workshop on Unified Theory and
Baryon number in the Universe}, eds. O. Sawada and A. Sugamoto (KEK 1979);
R.N. Mohapatra, and G. Senjanovi{\'c}, Phys. Rev. Lett. {\bf 44}, 912
(1980).

\bibitem{Allahverdi1}
R.~Allahverdi and A.~Mazumdar, Phys.\ Rev.\ D {\bf 67}, 023509 (2003).

\bibitem{Berezhiani}
Z.~Berezhiani, A.~Mazumdar and A.~Perez-Lorenzana,
Phys.\ Lett.\ B {\bf 518}, 282 (2001);
R.~Allahverdi, B.~Dutta and A.~Mazumdar, Phys.\ Rev.\ D {\bf 67}, 
123515 (2003). 

\bibitem{Brandenberger}
V.~F.~Mukhanov, H.~A.~Feldman and R.~H.~Brandenberger,
Phys.\ Rept.\  {\bf 215}, 203 (1992);



\bibitem{Liddle}
A.~D.~Linde and V.~Mukhanov, Phys.\ Rev.\ D {\bf 56}, 535 (1997);
A.~R.~Liddle and A.~Mazumdar, Phys.\ Rev.\ D {\bf 61}, 123507 (2000).

\bibitem{Ellis84}
J. Ellis, J E Kim, and D. V. Nanopoulos, Phys. Lett. B {\bf 145}, 181 (1984);

\bibitem{Maroto}
A. L. Maroto, and A. Mazumdar, Phys. Rev. Lett. {\bf 84},
1655 (2000).

\bibitem{WMAP}
H.~V.~Peiris {\it et al.},
arXiv:astro-ph/0302225.





\end{references}
\end{document}